\documentclass[aps,pra,twocolumn,superscriptaddress]{revtex4}
\usepackage[latin9]{inputenc}
\usepackage{amsmath}
\usepackage{amssymb}
\usepackage{esint}

\usepackage{epsfig}
\usepackage{bm}
\usepackage{amsfonts}
\usepackage{color}
\usepackage{bigints}
\usepackage{ulem}

\usepackage{amsmath}
\usepackage{latexsym}
\usepackage{mathrsfs}
\usepackage[sans]{dsfont}

\makeatletter

\newcommand {\scat} {\mathrm{sc}}

\newcommand {\ellg} {\ell_\mathrm{g}}

\newcommand {\D} {\mathrm{D}}
\newcommand {\dd} {\mathrm{d}}

\newcommand  {\g}   {\mathrm{g}}

\newcommand  {\crit}   {\mathrm{cr}}
\newcommand  {\spa}   {\mathrm{sp}}
\newcommand  {\tr}   {\mathrm{tr}}
\newcommand  {\INLN}   {Universit\'e C\^ote d'Azur, CNRS, INLN, Valbonne, France}

\begin{document}

\title{Diffusive to quasi-ballistic random laser: incoherent and coherent models}

\author{W. Guerin}
\email{william.guerin@inln.cnrs.fr}
\affiliation{\INLN}
\author{Y. D. Chong}
\affiliation{School of Physical and Mathematical Sciences, Nanyang Technological University, Singapore 637371, Singapore}
\author{Q. Baudouin}
\affiliation{\INLN}
\author{M. Liertzer}
\affiliation{Institute for Theoretical Physics, Vienna University of Technology (TU-Wien), A-1040 Vienna, Austria}
\author{S. Rotter}
\affiliation{Institute for Theoretical Physics, Vienna University of Technology (TU-Wien), A-1040 Vienna, Austria}
\author{R. Kaiser}
\affiliation{\INLN}

\date{\today}

\begin{abstract}
We study the crossover between the diffusive and quasi-ballistic regimes of random lasers. In particular, we compare incoherent models based on the diffusion equation and the radiative transfer equation (RTE), which neglect all wave effects, with a coherent wave model for the random laser threshold.
We show that both the incoherent and the coherent models predict qualitatively similar thresholds, with a smooth transition from a diffuse to a quasi-ballistic regime. The shape of the intensity distribution in the sample as predicted by the RTE model at threshold is also in good agreement with the coherent model. The approximate incoherent models thus provide useful analytical predictions for the threshold of random lasers as well as the shape of the random laser modes at threshold.
\end{abstract}

\maketitle

\section{Introduction}

Random lasers are probably among the most exotic sources of coherent light studied so far~\cite{Cao:2003,Cao:2005b,Wiersma:2008,Andreasen:2011,Wiersma:2013}. As their name already suggests, random lasers get their optical feedback not by external mirrors or through a resonator, but rather by the random scattering of light in a disordered medium. Even though this operational principle makes a deliberate tuning or selection of desired laser modes and output frequencies technically rather involved~\cite{Wiersma:2001,ElDardiry:2011,Leonetti:2013,Hisch:2013,Bachelard:2014}, first promising applications of random lasers are recently emerging for which these cost-efficient devices are ideally suited~\cite{Redding:2012,Schonhuber:2016}. From the fundamental point of view, random lasers offer an exciting research area at the interface between mesoscopic physics, non-Hermitian optics, and laser physics~\cite{Cao:2005b,Wiersma:2008,Rotter:2014}. Particularly exciting in this context is the hypothesis, first put forward by Lethokov~\cite{Letokhov:2009}, that random lasers may actually be occurring also on a natural basis in stellar gases, since both multiple scattering and amplification are present in such media~\cite{Strelnitski:1996,Johansson:2004,Johansson:2005}.
Such hitherto unobserved ``astrophysical random lasers'' would have a spatial extension many orders of magnitudes larger than the micron-sized random lasers that are meanwhile routinely fabricated in the laboratory~\cite{Cao:2003b}.

The vastly different length scales on which random lasing may occur, and the many different physical systems in which they have been realized, have triggered the development of different theoretical approaches to describe this phenomenon~\cite{Letokhov:1968,Wiersma:1996,Burin:2001,Pierrat:2007,Vanneste:2007,Conti:2008,Tureci:2008,Frank:2009}. Whereas it might appear reasonable that a radiative transfer approach, which does not incorporate interference effects,
may be appropriate for astronomical length scales with long amplifying paths and few scattering events, and a diffusive model may be suitable to describe strongly scattering media in the diffusive limit~\cite{Pinheiro:2006}, it has so far remained unexplored how to describe the crossover between such different regimes. An important aspect that is also missing in the literature is a global perspective on random lasing in which all the possible random lasing regimes are charted and properly identified.

The aim of this paper will be to take such a bird's eye perspective on random lasing and to connect different approaches with each other. In particular, we will focus on the general question which size a medium with a certain amount of gain and disorder needs to have such that it reaches the random lasing threshold. To address this problem we employ approximative tools such as the radiative transfer equation (RTE) (for the low-scattering, or quasi-ballistic, limit) as well as a diffusive model (for the strongly scattering limit) and compare them with a full solution of the scalar wave equation that encompasses both of these limits just as well as the crossover region in between. While the latter model includes diffraction and interference effects and relies on heavy numerical simulations, the diffusive and radiative transfer models are ``incoherent'' in the sense that they neglect all wave effects. Their advantage is to provide simple analytical results.

Our paper is organized as follows. In Sec. 2 we first briefly recall Letokhov's seminal results on the threshold of random lasers based on the diffusion equation~\cite{Letokhov:1968}. In Sec. 3 we present a theory of the random laser threshold based on the RTE. In Sec. 4 we introduce the employed coherent wave model. Finally, in Sec. 5 we compare the results from these different models and discuss the conclusions that can be drawn from them. A short summary is presented in Sec. 6.

\section{Random laser threshold from the diffusion equation}\label{sec.diffusion}

We summarize here the well-known results of Letokhov on the random laser threshold~\cite{Letokhov:1968}. The presentation is inspired by the one given in the review~\cite{Cao:2003}. We start from the diffusion equation for light with a gain term,
\begin{equation}\label{eq.diff_gain}
\frac{\partial W(\bm{r},t)}{\partial t} = D \nabla^2 W(\bm{r},t) + \frac{v_E}{\ellg} \; ,
\end{equation}
where $W$ is the energy density, $v_E$ is the energy transport velocity inside the medium, $\ell_\mathrm{g}$ is the gain length and $D$ is the diffusion coefficient. At 2D or 3D, it reads
\begin{equation}\label{eq.diff_coeff}
D_{2\D} = \frac{v_E\, \ell_\scat}{2} \, , \quad D_{3\D} = \frac{v_E\, \ell_\scat}{3} \, ,
\end{equation}
where $\ell_\scat$ is the mean free path. For simplicity, we consider only isotropic scatterers such that the mean free path is equal to the transport length~\cite{Cao:2003}. To map our results also to the case of finite-size scatterers, which scatter anisotropically, the scattering mean free path $\ell_\scat$ needs to be rescaled to the transport length $\ell_\tr = \ell_\scat/(1- \cos \phi)$, where $\phi$ is the scattering angle. The scatterers need to stay below the wavelength though, as shape-specific resonances would otherwise destroy the universality of our analysis~\cite{Garcia:2007}.

Using the modal decomposition
\begin{equation}\label{eq.diff_gain}
W(\bm{r},t) = \sum_n a_n \Psi_n(\bm{r}) e^{(D B_n^2 - v_E/\ell_\g) t} \; ,
\end{equation}
with appropriate boundary conditions, one can show that the threshold of a random laser is reached when
\begin{equation}
D B_1^2-\frac{v_E}{\ell_g} = 0 \; ,
\end{equation}
where $B_1$ is the smallest eigenvalue, corresponding with the longest-lived mode. For a 3D sphere of radius $R$, $B_1 = \pi/R$ and for a 2D disk of radius $R$, $B_1 = j_{0,0}/R$, where $j_{0,0}\simeq 2.40$ is the first root of the Bessel function $J_0$.

Finally it leads to the following critical radius
\begin{equation}\label{eq.Lcr}
R_\crit^{3\D} = \pi \sqrt{\frac{\ell_\scat\,\ell_\mathrm{g}}{3}} \, ,  \quad R_\crit^{2\D} = 2.40 \sqrt{\frac{\ell_\scat\,
\ell_\mathrm{g}}{2}} \, .
\end{equation}
Note that the numerical factors in front of $(\ell_\scat \ell_\g)^{1/2}$ differ from each other by only a few percents. Note also that we have neglected here the ``extrapolation length''~\cite{Rossum:1999,Ishimaru:book,note_extrapolation_length}, which is a small correction in the diffusive limit that we consider in this section.
The diffusive, or multiple-scattering regime, is reached when $R \gg \ell_\scat$, which corresponds to the validity range of this threshold condition.

\section{Random laser threshold from the radiative transfer equation}\label{sec.RTE}

In a regime of low scattering, transport of light is no longer governed by a diffusive equation, but is well described by the radiative transfer equation (RTE). The RTE is used in many different fields dealing with transport in complex media, such as astrophysics~\cite{Chandrasekhar:1960,Sobolev:1963,Peraiah:2002}, neutron physics~\cite{Zweifel}, or biological imaging~\cite{Wang:book}. The diffusion equation can be derived from the RTE with supplementary approximations (see, e.g., Refs.~\cite{Ishimaru:book,Wang:book}). The RTE is thus more general and has been shown to be valid from the ballistic regime to the diffusive one~\cite{Elaloufi:2004}. It neglects, however, all wave effects like interference and diffraction.

The basic quantity of the RTE is the ``radiance'' or ``specific intensity'' $L(\bm{r},\bm{u},t)$, which describes the photon density at point $\bm{r}$, propagating along direction $\bm{u}$ at time $t$.
In a system exhibiting absorption and scattering, the RTE reads
\begin{equation}
\begin{split}
 \frac{1}{c} \, \frac{\partial L}{\partial t}(\bm{r},\bm{u},t)+ \bm{u} \cdot & \bm{\nabla} L(\bm{r},\bm{u},t) = - (\alpha+\chi) L(\bm{r},\bm{u},t) \\
 & +  \frac{\chi}{4\pi} \int_{0}^{4\pi}  p(\bm{u},\bm{v}) L(\bm{r},\bm{v},t) \, \mathrm{d}\Omega \; ,
\end{split}\label{eq:ETR1}
\end{equation}
where $\alpha$ is the linear absorption coefficient, $\chi = \ell_\scat^{-1}$ and $p(\bm{u},\bm{v})$ describes the scattering angular diagram. For a medium with gain, $\alpha<0$ and we can also use the linear gain coefficient $g=-\alpha = \ell_\g^{-1} > 0$. The RTE can be derived from Maxwell equations~\cite{Pierrat:thesis} but can also be found by simple energy conservation arguments, since it is a Boltzmann-type equation.

From the specific intensity one can define two other useful quantities, the radiative flux $\bm{q}(\bm{r},t)$, which is identical to the Poynting vector, and the energy density $W(\bm{r}, t)$, which is the quantity entering into the diffusion equation:
\begin{gather}
\bm{q}(\bm{r},t) = \int_{4\pi} L(\bm{r},\bm{v},t) \bm{u} d\Omega\; , \label{eq.flux_def}\\
W(\bm{r}, t) = \int_{4\pi} \frac{L(\bm{r},\bm{v},t)}{c} d\Omega\; .\label{eq.Energydensity_def}
\end{gather}


\subsection{Random laser threshold}

For a slab geometry, the random laser threshold was found from the RTE using a modal decomposition~\cite{Pierrat:2007}, and applied to the case of a random laser based on cold atoms~\cite{Froufe:2009,Guerin:2010}.
For a sphere geometry, Letokhov \textit{et al.} have also derived the random laser threshold from the RTE~\cite{Lavrinovich:1975,Johansson:2007,Letokhov:2009}. The detailed derivation can be found in Ref.~\cite{Quentin:thesis}; we only recapitulate the result here. Moreover, for a better comparison with the data obtained from the coherent wave model (Sec.~\ref{sec.SALT}), we have extended these results to the case of a 2D disk. We also give only the result in this section; a detailed derivation is provided in Appendix~A.

For a 3D sphere one obtains a critical radius for the random laser threshold $R_\crit$ given by~\cite{Lavrinovich:1975,Johansson:2007,Letokhov:2009,Quentin:thesis}
\begin{equation}\label{eq.Letokhov_generalise}
\tan(qR_\crit) = \frac{2gqR_\crit}{2g-q^2R_\crit} \; ,
\end{equation}
with
\begin{equation}\label{eq.q}
q^2 = 3g(\chi-g) = \frac{3}{\ell_\g}\left(\frac{1}{\ell_\scat}-\frac{1}{\ell_\g}\right) \, .
\end{equation}

For a 2D disk, the threshold condition is
\begin{equation}\label{eq.threshold_finale}
\frac{J_0(\beta R_\crit)}{J_1(\beta R_\crit)} = \frac{\pi}{2} \frac{g}{\beta} \, ,
\end{equation}
where $J_0$, $J_1$ are Bessel functions of the first kind, and with
\begin{equation}\label{eq.beta}
\beta^2 = 2g(\chi-g)= \frac{2}{\ell_\g}\left(\frac{1}{\ell_\scat}-\frac{1}{\ell_\g}\right) \, .
\end{equation}

\begin{figure}[tb]
\centering
\includegraphics{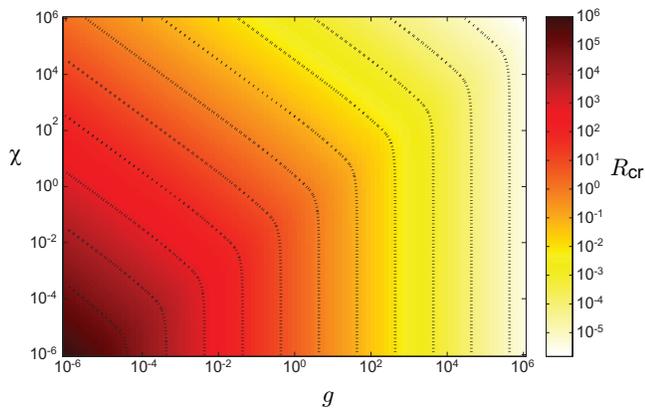}
\caption{Critical radius for the random laser threshold as a function of the gain coefficient $g=\ell_\g^{-1}$ and the scattering coefficient $\chi = \ell_\scat^{-1}$, as given by the numerical solution of Eq.~\ref{eq.Letokhov_generalise}. Note the log scales. The dotted lines are iso-$R_\crit$ contours.}
\label{fig.Letokhov_generalise}
\end{figure}

These threshold equations can easily be solved numerically and give results which are very close to each other. We show in Fig.~\ref{fig.Letokhov_generalise} the result for the 3D case.
It shows a smooth transition from the diffusive regime (upper-left part) to the quasi-ballistic regime (lower-right part).

\subsection{Limiting cases}

\subsubsection{Diffusive limit}

For the 3D case, one can recover the diffusive threshold given by Eq.~(\ref{eq.Lcr}) from Eqs.~(\ref{eq.Letokhov_generalise}-\ref{eq.q}) by supposing that there is much more scattering than gain, $\chi \gg g$, so that $q^2 \simeq 3g\chi$, and also that $\chi R_\crit \gg 1$ (diffusive regime). One can then easily show that the r.h.s. of Eq.~(\ref{eq.Letokhov_generalise}) is very small. Then  Eq.~(\ref{eq.Letokhov_generalise}) simplifies to $qR_\crit \sim \pi$, which gives $R_\crit \sim \pi/q \sim \pi (\ell_\g \ell_\scat/3)^{1/2}$ as expected.

For the 2D case, supposing also that $\chi \gg g$, then $\beta^2 \simeq 2g\chi$, and the threshold equation reduces to
\begin{equation}
\frac{J_0 (\beta R_\crit)}{J_1 (\beta R_\crit)} = \frac{\pi g}{2\beta} \simeq \frac{\pi}{2}\sqrt{\frac{g}{2\chi}} \ll 1 \; .
\end{equation}
We can thus take the zero of the function $J_0(z)/J_1(z)$, which is the zero of $J_0(z)$, i.e., $j_{0,0}\simeq 2.40$. Thus $\beta R_\crit \sim 2.40$ and we recover $R_\crit \sim 2.40 (\ell_\g \ell_\scat/2)^{1/2}$.

\subsubsection{Ballistic limit}

Interestingly, one can also simplify the threshold equations in the opposite limit of very low scattering and high gain..

For the 3D case, if $\chi \ll g$, we have the simplification $q\simeq \pm ig\sqrt{3}$ and $\tan(qR_\crit) \simeq \tan(\pm i \sqrt{3}gR_\crit)$. For $g R_\crit > 1$ it gives $\tan(qR_\crit)\simeq \pm i$. Then Eq.~(\ref{eq.Letokhov_generalise}) is easily solved to~\cite{Note_misprint}.
\begin{equation}\label{eq.ballistic3D}
R_\crit \sim \frac{1}{\left(\sqrt{3}-3/2\right)g} \approx 4.31 \ell_\g \, .
\end{equation}

At 2D, if $\chi \ll g$, $\beta^2 \simeq -2 g^2$, $\beta \simeq \pm i \sqrt{2} g$, and the threshold equation reduces to
\begin{equation}
\frac{\pm i \sqrt{2}\, J_0 (\pm i \sqrt{2} g R_\crit)}{J_1 (\pm i \sqrt{2} g R_\crit)} = \frac{\pi}{2} \, .
\end{equation}
The solution is
\begin{equation}\label{eq.ballistic2D}
R_\crit \approx 3.76 \ell_\g \, .
\end{equation}

In both cases, we obtain a finite critical radius that does not depend on the scattering $\chi$, corresponding to the vertical asymptotes in Fig.~\ref{fig.Letokhov_generalise}. Very surprisingly, this result suggests that a threshold exists even without scattering, a conclusion that seems clearly unphysical, suggesting that some of the approximations made to derive the RTE threshold (see Appendix A) break down in the ballistic limit. We discuss in more detail the nature of the employed approximations in Appendix B.

\subsection{Shape of the energy density at threshold}

The shape of the intensity distribution at threshold can be obtained by solving Eq.~(\ref{eq.finale}) (or its 3D equivalent).

In 3D one finds~\cite{Quentin:thesis}
\begin{equation}\label{eq.intensity_profile3D}
W_0(\bm{r}) \propto \frac{\sin(qr)}{r} \, ,
\end{equation}
while in 2D we obtain
\begin{equation}\label{eq.intensity_profile2D}
W_0(\bm{r}) \propto J_0(\beta r) \, .
\end{equation}

In both cases, if there is more scattering than gain, $\chi > g$, $\beta$ and $q$ are real and $W_0(\bm{r})$ is bell-shaped with its maximum at $r=0$. On the contrary, if $\chi < g$, $\beta$ and $q$ are purely imaginary and $W_0(\bm{r})$ increases from the center (Fig.~\ref{fig.Comparison_shape}). This is consistent with what could be expected in a quasi-ballistic regime, where photons farther from the center have been in averaged more amplified.

\section{Random laser threshold from coherent wave calculations}
\label{sec.SALT}

In order to compare the predictions of the RTE with a more complete model, we use
coherent wave calculations of the lasing threshold, which account for
the effects of finite wavelengths and wave interference.  Due to the
computational difficulty of performing such calculations on disordered
media, we restrict the comparison study to 2D, using the scalar wave
equation
\begin{equation}
  \left[\nabla^2 + \epsilon(\bm{r},\omega)
    \left(\frac{\omega}{c}\right)^2\right] \psi(\bm{r}) = 0.
  \label{wave equation}
\end{equation}
This describes a 2D electromagnetic mode in the transverse magnetic
(TM) polarization, where $\psi(\bm{r})$ is the complex scalar
wavefunction corresponding to the out-of-plane component of the
electric field, $\omega$ is the mode frequency, $\nabla^2$ is the 2D
Laplacian, and $\epsilon(\bm{r},\omega)$ is the dielectric function.

The wave equation (\ref{wave equation}) introduces an extra length
scale, the wavelength $\lambda \sim 2\pi c/\omega$.  For comparison to
the RTE, we shall be interested in the regime where the wavelength is
shorter than the other length scales, i.e.,~$c/\omega \ll \{R,
\ell_\scat, \ell_g\}$.

We model the random laser by uniformly distributing $N$ delta-function
scatterers at positions $\left\{\bm{r}_1,\dots,\bm{r}_N\right\}$,
within a circular region of radius $R$.  This region also contains a
uniform background of gain material, with susceptibility $\chi_g \in
\mathbb{C}$.  Thus,
\begin{equation}
  \epsilon(\bm{r},\omega) =
  \left\{\begin{array}{ll}
  1 + \chi_g + a \sum_{j=1}^N \delta^2(\bm{r}-\bm{r}_j),& \;\;\;r \le R\\
  1, & \;\;\;r > R.
  \end{array} \right.
  \label{epsn}
\end{equation}
The parameter $a$, which has units of area, determines the strength of
each scatterer.  The use of independent delta-function scatterers
allows us to relate the model parameters to the mean free path.  The
density of scatterers is $\rho = N / \pi R^2$, and the 2D scattering
cross section of an individual scatterer in the first Born
approximation is $\sigma = a^2(\omega/c)^3/4$.  Thus,
\begin{equation}
  \ell_\scat = \frac{1}{\rho\sigma} = \frac{4\pi R^2}{N a^2 (\omega / c)^3}.
  \label{ells}
\end{equation}

From this setup, the lasing threshold calculation proceeds as follows:
for a fixed lasing frequency, scatterer distribution, and scatterer
strength, we find a complex value of $\chi_g$ that satisfies
Eq.~(\ref{wave equation}) with purely-outgoing boundary conditions.
The detailed procedure is described in Appendix C.  Essentially, we
perform a partial-wave expansion on $\psi(\bm{r})$, which reduces
Eq.~(\ref{wave equation}) to a non-Hermitian eigenproblem whose
eigenvalues are the values of $\chi_g$ for which the solution is
purely-outgoing in the external region $r > R$.  Out of these possible
values of $\chi_g$, we choose one with sufficiently small
$\mathrm{Re}[\chi_g]$ (i.e., negligible index shift), and the smallest
value of $-\mathrm{Im}[\chi_g]$ (i.e., least gain needed to reach
threshold).  This mode's refractive index is
\begin{equation}
  n_g \; \approx\; 1 + \frac{i}{2}\, \Big(\mathrm{Im}[\chi_g]\Big).
\end{equation}
By repeating this procedure for many realizations of the scatterer
distribution, we compute
\begin{equation}
 g = \ell_g^{-1} \,\equiv\, \Big\langle -2\mathrm{Im}[n_g] \omega/c
  \Big\rangle
  \,=\, \frac{\omega}{c} \Big\langle
  -\mathrm{Im}[\chi_g] \Big\rangle.
  \label{ellg}
\end{equation}
As in the RTE, $\ell_g$ represents the average path length traveled by
a photon before an amplification event.  By changing the individual
scatterer strength $a$ and using Eq.~(\ref{ells}), we can find the
dependence of $\ell_g$ on $\ell_\scat$, and compare the result to the
predictions of the RTE.

We perform two sets of calculations, for $\omega = 30 c/R$ and $\omega =
60 c/R$; as we shall see, these two frequencies give qualitatively
similar results.  For each case, we take $N = 250$ scatterers, and
tune $a$ so that the mean free path varies over $10^{-1}R \lesssim
\ell_\scat \lesssim 10^2R$, ranging from the diffusive to the
quasi-ballistic regime.

\section{Comparison}\label{sec.comparison}

In this section we compare the results of the different models for the threshold and for the intensity distribution at threshold.

\subsection{Threshold}

\begin{figure}[t]
\centering
\includegraphics{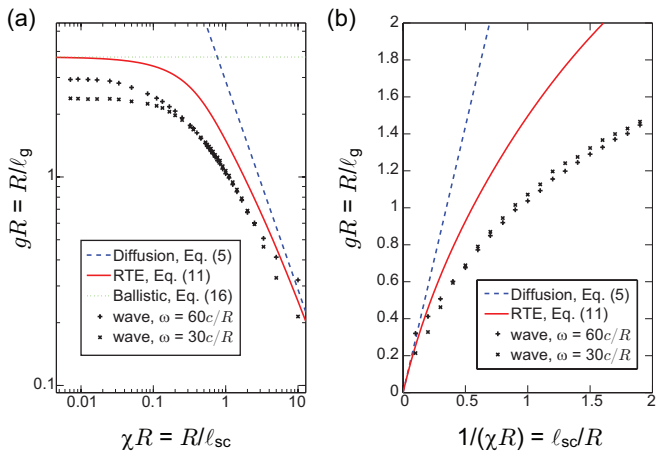}
\caption{(a) Comparison of the thresholds computed from the different models. Points are numerical solutions of the wave model for two different frequencies. The dashed blue line is the diffusive threshold computed from Eq.~(\ref{eq.Lcr}). The red solid line is the RTE threshold computed from Eq.~(\ref{eq.threshold_finale}). Its asymptotic behavior in the ballistic regime (Eq.~\ref{eq.ballistic2D}) is indicated by the green dotted line. Note the logarithmic scales. (b) Zoom into the diffusive and intermediate regimes (linear scales and inverted x-axis).}
\label{fig.Comparison2D}
\end{figure}

We use the different models to plot $gR$ at threshold as a function of $\chi R = R/\ell_\scat$ or $1/(\chi R)$. We show in Fig.~\ref{fig.Comparison2D} the comparison between the data of the coherent wave model (previous section) and the analytical results of the diffusion (sec.~\ref{sec.diffusion}) and RTE thresholds (sec.~\ref{sec.RTE}).



Overall, looking at Fig.~\ref{fig.Comparison2D}(a), we can observe that the wave-model and the RTE thresholds are quite close to each other. Moreover, the wave-model threshold becomes very close to the diffusive ones for large optical thickness $\chi R$. The fact that fully incoherent models provide here very good estimates was not obvious from the outset because the incoherent models are only expected to describe transport properties averaged over the disorder configurations. On the contrary, the coherent model selects the ``best'' mode at each realization (see Fig.~\ref{fig.chig}). This also explains why the incoherent models predict larger gain thresholds and are thus ``pessimistic''. The discrepancy between the different models increases as the optical thickness ($\propto \chi R$) decreases (see Fig.~\ref{fig.Comparison2D}b).

Another important observation is that the RTE threshold is significantly more accurate (closer to the wave model) than the threshold from the diffusion equation. For example, in the intermediate regime $R \approx \ell_\scat$, the wave model predicts a gain threshold $gR \approx 1.1$, while the RTE threshold is $gR \approx 1.5$ and the diffusive one is $gR \approx 2.9$. Thus, as soon as the random laser is not deeply in the diffusive regime, the RTE theory provides a significant improvement.


However, in the limit of very low scattering (ballistic, or empty disk), the RTE model predicts a scattering-independent finite threshold. As already mentioned, this indicates a breakdown of the approximations used to derive the threshold in the RTE model. On the contrary, the scattering-independent threshold of the coherent model can have a clear physical interpretation: the disk boundary creates an index mismatch with the surrounding vacuum due to the gain coefficient, and thus reflects some light, which induces some coherent feedback. In this regime, the laser is not ``random'' and is based on whispering gallery modes. The corresponding gain threshold depends on the wavelength because the index mismatch depends on the wavelength for a fixed gain coefficient. The coherent model is thus able to describe the transition from a diffusive random laser to a ``ballistic'', cavity-based one. The partial reflection due to the index mismatch is not included in the incoherent models.


\subsection{Intensity distribution}

\begin{figure}[t]
\centering
\includegraphics{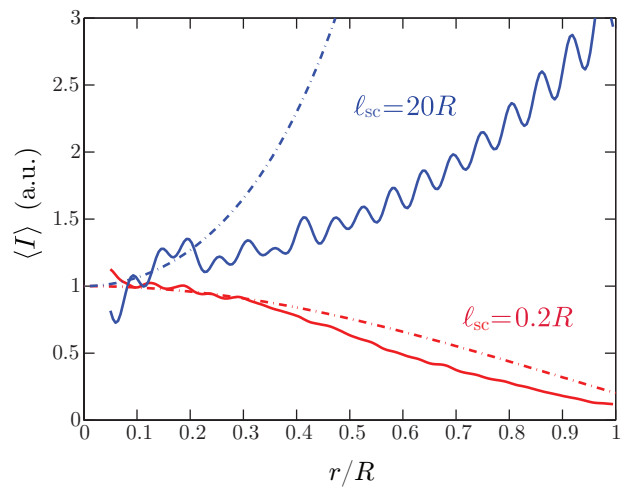}
\caption{Intensity distribution at threshold averaged over the disorder and over the radial angle. The solid lines are computed with the coherent wave model and the dash-dotted lines with the analytical results of the RTE model [Eq.~(\ref{eq.intensity_profile2D})]. The vertical scale has been chosen such that $\langle I \rangle \sim 1$ at the center. In the diffusive regime ($\ell_\scat = 0.2 R$), the energy is confined near the center, while in the quasiballistic regime ($\ell_\scat = 20 R$) it increases from the center toward the edge.}
\label{fig.Comparison_shape}
\end{figure}

We also can compare the averaged intensity distribution of the lasing mode at threshold obtained from the wave model and the analytical profile predicted by the RTE model (see Eq.~\ref{eq.intensity_profile2D}). We show in Fig.~\ref{fig.Comparison_shape} the intensity profile computed in the wave model at threshold for $\omega =60 c/R$, averaged over 100 disorder configurations and over the radial angle, for two different scatterers strengths (solid lines). For highly scattering samples ($\ell_\scat = 0.2 R$), we observe that the energy is confined near the center, as it could be expected in the diffusive regime. On the contrary, for weakly disordered samples ($\ell_\scat = 20 R$), the averaged intensity increases from the center (this can also be see in the Fig.~4 of Ref.~\cite{Tureci:2008}). These qualitatively different behaviors are well captured by the RTE prediction (dash-dotted lines). The agreement is even quite good in the diffusive regime. Differences are more important in the quasiballistic regime. First, the gain at threshold is higher in the RTE model [Fig.~\ref{fig.Comparison2D}] and thus the intensity increases faster than in the wave model. Second, oscillations appear in the coherent model, which are a signature of interference effects due to the partial reflection at the boundary, creating an oscillatory pattern in the lasing mode. This partial reflection also contributes to increasing the intensity at the center, reducing the difference between the center and the edge.

\section{Conclusion}

In this article, we have studied the threshold of random lasing in a 2D disk in the crossover from the diffusive to the quasi-ballistic regimes, and we compared different models to describe this transition. The more accurate, coherent model is able to describe this system in all regimes, at the cost of computational complexity, which, in particular, limits the size and dimensionality of the system under study. In the diffusive limit, the diffusion equation, which provides a fully incoherent description, predicts the threshold quite accurately, although interference and wave effects are neglected. Moreover, an incoherent model is also available beyond the diffusive regime.
This model, based on the radiative transfer equation, also provides analytical results, which agree with the full coherent model on a qualitative level, and even show quantitative agreement in the crossover regime on a level superior to the diffusive model. The radiative transfer equation also correctly predicts the global shape of the averaged intensity distribution at threshold.

Surprisingly, even though the incoherent model is expected to break down deep in the ballistic limit because the boundary conditions cannot be treated rigorously, it predicts a random laser threshold as well as a modal intensity distribution in qualitative agreement with the coherent model.
At the other extreme, the diffusion model should also break down when the mean-free path becomes comparable to the wavelength.
Despite these limitations, the incoherent models are very efficient in predicting the good order of magnitude for the random laser threshold in a large range of parameters.

The comparison with experimental data would also be very interesting. In most experimentally accessible systems, however, the polydispersity of the samples and the complicated geometry due to the scattering of the pump~\cite{Noginov:2006} makes a quantitative comparison very hard. These problems are reduced with cold atoms and the observed threshold reported in Ref.~\cite{Baudouin:2013b} was not very far from Letokhov's diffusive threshold, showing that such simplified models can be useful guides to experimentalists.


\ \\

\noindent\textbf{Funding.} Agence National pour la Recherche (ANR, project CAROL, No. ANR-06-BLAN-0096); D\'el\'egation G\'en\'erale \`a l'Armement (DGA); Singapore National Research Foundation
(grant No.~NRFF2012-02); Singapore MOE Academic Research Fund Tier 3 (grant MOE2011-T3-1-005); Austrian Science Fund (FWF, projects No. SFB-NextLite F49-P10 and No. I 1142-N27 GePartWave) .

\ \\

\noindent\textbf{Acknowledgment.} We acknowledge A.~D.~Stone for helpful discussions and guidance.


\section*{APPENDIX A: Derivation of the RTE threshold at 2D}

We present here a detailed derivation of the random laser threshold (Eq.~\ref{eq.threshold_finale}), as well as the shape of the intensity distribution at threshold (Eq.~\ref{eq.intensity_profile2D}), from the radiative transfer equation for a 2D disk.

\subsection*{Threshold condition from the RTE}

We start from the RTE written at 2D,
\begin{equation}
\begin{split}
 \frac{1}{c} \, \frac{\partial L}{\partial t}(\bm{r},\bm{u},t)+ \bm{u} \cdot \bm{\nabla} L(\bm{r},\bm{u},t) = - (\alpha+\chi) L(\bm{r},\bm{u},t)  + \\
  \frac{\chi}{2\pi} \int_{0}^{2\pi}  p(\bm{u},\bm{v}) L(\bm{r},\bm{v},t) \, \mathrm{d}\theta' \; ,
 \end{split}
    \label{eq:ETR2}
\end{equation}
where $\theta' = (\bm{r},\bm{v})$ is the plane angle between $\bm{r}$ and $\bm{v}$.
In the following, we suppose isotropic scattering: $p(\bm{u},\bm{v})=1$. Then the last integral reads $\int  L(\bm{r},\theta,t) \mathrm{d}\theta$ where $\theta$ is in the following the angle between $\bm{r}$ and $\bm{u}$.
In cylindrical coordinate the gradient is
\begin{equation}
\bm{\nabla} L = \frac{\partial L}{\partial r} \bm{e_r} + \frac{1}{r} \frac{\partial L}{\partial \theta} \bm{e_\theta}
\end{equation}
and we have $\bm{u} \cdot \bm{e_r} = \cos \theta$ and $\bm{u} \cdot \bm{e_\theta} = -\sin \theta$.
We thus obtain
\begin{equation}
\frac{1}{c} \, \frac{\partial L}{\partial t}+ \cos\theta \frac{\partial L}{\partial r} - \frac{\sin \theta}{r} \frac{\partial L}{\partial \theta} = - (\alpha+\chi) L +  \frac{\chi}{2\pi} \int_{0}^{2\pi}  L \, \mathrm{d}\theta .
    \label{eq.ETR3}
\end{equation}

We now look for a separable solution in the form
\begin{equation}\label{eq.separation}
L(\bm{r}, \theta, t) = L_t(t) \times L_\spa (\bm{r},\theta) \, ,
\end{equation}
where ``sp'' means ``space''. Injecting Eq.~(\ref{eq.separation}) into Eq.~(\ref{eq.ETR3}) we obtain
\begin{equation}
\begin{split}
& \frac{1}{c} \, \frac{\partial L_t}{\partial t} = \frac{L_t}{L_\spa} \\
& \times \left[ - \cos\theta \frac{\partial L_\spa}{\partial r} + \frac{\sin \theta}{r} \frac{\partial L_\spa}{\partial \theta} - (\alpha+\chi) L_\spa +  \frac{\chi}{2\pi} \int_{0}^{2\pi}  L_\spa \, \mathrm{d}\theta \right] \, .
 \end{split}
\end{equation}
This is an equation in the form $\partial L_t / \partial t = c S L_t$, which induces an exponential increase when $S > 0$. The threshold condition is thus $S=0$, i.e.,
\begin{equation}\label{eq.threshold_RTE}
\cos\theta \frac{\partial L_\spa}{\partial r} - \frac{\sin \theta}{r} \frac{\partial L_\spa}{\partial \theta} = -(\alpha+\chi) L_\spa +  \frac{\chi}{2\pi} \int_{0}^{2\pi}  L_\spa \, \mathrm{d}\theta \, .
\end{equation}

\subsection*{The Eddington approximation}

Unfortunately this equation is still hard to solve and we need an approximation. Following Letokhov \textit{et al.}~\cite{Lavrinovich:1975,Johansson:2007,Letokhov:2009}, we use the derivation of Sobolev~\cite{Sobolev:1963} based on the so-called Eddington approximation~\cite{Unno:1966,Wiscombe:1977}. It consists in supposing that the second moment of the luminance $L_\spa$ respective to the cosine of the propagation angle is proportional to the zeroth one. It is equivalent to writing
\begin{equation}\label{eq.Eddington}
L_\spa (r, \theta) = a(r) + b(r) \cos(\theta) \, .
\end{equation}
We can then derive several useful relations:
\begin{align}
L_0(\bm{r}) &= \frac{1}{2\pi} \int_0^{2\pi} L_\spa (\bm{r},\theta) \mathrm{d}\theta &=& \, a(\bm{r}) \, ,\label{eq.L0}\\
L_1(\bm{r}) &=\frac{1}{2\pi} \int_0^{2\pi} L_\spa (\bm{r},\theta) \cos\theta \mathrm{d}\theta &=& \, \frac{b(\bm{r})}{2} \, ,\label{eq.L1}\\
L_2(\bm{r}) &= \frac{1}{2\pi} \int_0^{2\pi} L_\spa (\bm{r},\theta) \cos^2\theta \mathrm{d}\theta &=& \, \frac{a(\bm{r})}{2} = \frac{L_0(\bm{r})}{2} \, ,\label{eq.L2}\\
\frac{\partial L_\spa}{\partial\theta} &= -\sin(\theta)\, b(\bm{r}) \, .&\label{eq.dLdtheta}
\end{align}

We can use these relations to simplify the $\theta$ dependency in the threshold equation (\ref{eq.threshold_RTE}).
For this, we first integrate Eq.~(\ref{eq.threshold_RTE}) over $\theta$ and, using
Eq.~(\ref{eq.L1}) and Eq.~(\ref{eq.dLdtheta}), we obtain
\begin{equation}\label{eq.ODE_L1}
\frac{\dd L_1}{\dd r} + \frac{L_1}{r} = -\alpha L_0 \, .
\end{equation}
Then we integrate again Eq.~(\ref{eq.threshold_RTE}) over $\theta$ after multiplication by $\cos(\theta)$
and we obtain
\begin{equation}\label{eq.ODE_L2}
\frac{\dd L_2}{\dd r} = -(\alpha+\chi) L_1 \, .
\end{equation}
Next we multiply Eq.~(\ref{eq.ODE_L1}) by $-(\alpha+\chi)$ and use Eq.~(\ref{eq.ODE_L2}) to obtain
\begin{equation}
\frac{\dd^2L_2}{\dd r} + \frac{1}{r} \frac{\dd L_2}{\dd r} = \alpha (\alpha+\chi) L_0 \, .
\end{equation}
Finally, since $L_2 = L_0/2$ [Eq.~(\ref{eq.L2})],
\begin{equation}\label{eq.finale}
\frac{\dd^2L_0}{\dd r} + \frac{1}{r} \frac{\dd L_0}{\dd r} = 2 \alpha (\alpha+\chi) L_0 \, .
\end{equation}

At this stage, we get a single differential equation on the quantity $L_0(\bm{r})$, which is the intensity distribution, with only one variable. Note that in the derivation at 3D, small differences appear because we integrate each time over the full solid angle, which makes a supplementary $\sin(\theta)$ appear in the integrals. We get at the end a very similar equation, with the factor $1/r$ replaced by $2/r$ and the 2 in the r.h.s. replaced by 3.

Another expression that will be useful in the following is obtained by combining Eqs.~(\ref{eq.L0}, \ref{eq.L2}, \ref{eq.ODE_L2}) into Eq.~(\ref{eq.Eddington}):
\begin{equation}\label{eq.Lsp}
L_\spa(\bm{r}, \theta) = L_0(\bm{r}) - \frac{1}{\alpha+\chi}\, \frac{dL_0}{dr} \cos(\theta) \, .
\end{equation}

\subsection*{Shape of the mode}

If we solve Eq.~(\ref{eq.finale}) we get the shape of the intensity distribution $L_0(\bm{r})$ at threshold. The solution of Eq.~(\ref{eq.finale}) that has no divergence at $r=0$ is
\begin{equation}\label{eq.intensity_profile}
L_0 (\bm{r}) = C J_0(\beta r) \, ,
\end{equation}
with $\beta^2 = -2\alpha(\alpha+\chi) = 2g(\chi-g)$, where $g=-\alpha$ is the gain coefficient, and $J_0$ is the Bessel function of the first kind of order 0.


\subsection*{Boundary conditions}

Since the random laser threshold obviously depends on the size of the medium, it comes from the boundary condition that should be applied to Eq.~(\ref{eq.finale}).

The medium has a finite radius $R$. The physical boundary condition should be that there is no ingoing intensity, i.e., $L_\spa(R,\theta) = 0$ for all $\theta$ such that $\cos\theta<0$. However, it is not possible to fulfill this condition consistently with the Eddington approximation (\ref{eq.Eddington}) (except for the trivial case of $L_\spa = 0$ everywhere). We thus have to use an approximate boundary conditions, which is that the total ingoing flux is zero:
\begin{equation}\label{eq.flux}
\int_{\cos\theta<0} L_\spa(R,\theta) \cos(\theta) \dd\theta \, = 0 .
\end{equation}
Note that the same problem appears with the use of the diffusion equation and the same approximate condition is used, leading to the extrapolation length (see, e.g., Ref.~\cite{Ishimaru:book}, p.~179).

We thus apply Eq.~(\ref{eq.flux}) to Eq.~(\ref{eq.Lsp}) to obtain $-2L_0(R) + \pi L_1(R) = 0$. Using Eqs.~(\ref{eq.L2},\ref{eq.ODE_L2}),
\begin{equation}
L_1(R) = -\frac{1}{\alpha+\chi}\, \frac{\dd L_2}{\dd r}|_R = -\frac{1}{\alpha+\chi} \, \frac{1}{2} \, \frac{\dd L_0}{\dd r}|_R \, ,
\end{equation}
and we obtain the approximate boundary condition
\begin{equation}\label{eq.boundary}
L_0(R) = -\frac{\pi}{4} \frac{1}{\alpha+\chi} \frac{\dd L_0}{\dd r}|_R \, .
\end{equation}
Note that the boundary condition for the 3D case is similar, the factor $\pi/4$ being replaced by $2/3$ (and is the same as in the diffusion approximation).

Using the intensity profile (\ref{eq.intensity_profile}) we finally get a threshold condition
\begin{equation}
J_0(\beta R) = \frac{\pi}{4} \frac{\beta}{\alpha+\chi} J_1(\beta R) \, .
\end{equation}
We can simplify $\beta/(\chi-g) = 2g/\beta$ and, since we use quantities that are normalized to the medium size, it is better to write the threshold condition in the following way:
\begin{equation}
\frac{\beta R \, J_0(\beta R)}{J_1(\beta R)} = \frac{\pi}{2} g R \qquad \mathrm{with} \qquad \beta^2 = 2g(\chi-g) \, .
\end{equation}

\section*{APPENDIX B: Discussion on the approximations used in the incoherent models}\label{sec.discussion}

From the initial RTE, there are several ways of finding a random laser threshold.

The first one is to derive first the diffusion equation, and then investigate the random laser problem. This approach gives the well-known results of Sec.~\ref{sec.diffusion}. The derivation of the diffusion equation from the RTE needs two approximations (see, e.g., Ref.~\cite{Wang:book} for a complete derivation, and Ref.~\cite{Olson:2000} for a discussion on various possible approximations). The first one is called the $P_1$ approximation, it consists in decomposing the specific intensity on the basis of Legendre polynomials of $\cos\theta$ and keeping only the first order. At 3D, using the definitions of the radiative energy density and flux [Eqs.~(\ref{eq.flux_def},\ref{eq.Energydensity_def})], it reads
\begin{equation}\label{eq.P1}
L(\bm{r}, \bm{u}, t) = \frac{c}{4\pi} W(\bm{r},t) + \frac{3}{4\pi} \bm{q}(\bm{r},t) \cdot \bm{u} \; .
\end{equation}
This approximation is good if the radiation is ``nearly'' isotropic. For this, photons need enough scattering events to randomize their directions, i.e., one needs $R \gg \ell_\scat$ and, in case of absorption, $\ell_\mathrm{a} \gg \ell_\scat$. The second approximation consists in neglecting the time derivative of the flux compared to the time scale associated to transport. This condition is usually said to be fulfilled if $\ell_\mathrm{a} \gg \ell_\scat$, where only absorption is considered. However, in case of gain, it adds the condition $\ell_\g \gg \ell_\scat$, which may be a limitation for the random laser problem, because it excludes the regime of parameters where there is more gain than scattering. Finally, to determine the random laser threshold from the diffusion equation, the boundary conditions due to the finite size of the medium are treated with an approximation that makes the extrapolation length appear~\cite{Rossum:1999,Ishimaru:book,note_extrapolation_length}. Because of these approximate boundary conditions, the diffusion equation is known to be bad near the borders of the medium (meaning at a few $\ell_\scat$).

In the approach presented in Sec.~\ref{sec.RTE}, we first find a complicated threshold equation directly from the RTE, and then, on this threshold equation, we make approximations. The Eddington approximation (Eq.~\ref{eq.Eddington}) is exactly the same as the $P_1$ approximation [note the similarity between Eqs.~(\ref{eq.P1}) and (\ref{eq.Lsp})], and the approximated boundary conditions (Eq.~\ref{eq.flux}) are also exactly the same as those used with the diffusion equation~\cite{Rossum:1999,Ishimaru:book}. The only condition that is relaxed is the one about the derivative of the flux. It relaxes the condition $\ellg \gg \ell_\scat$, which increases the validity range of the threshold condition to the case where the gain is similar or larger than the scattering. It is thus a significant improvement over the traditional Letokhov's threshold. However, the condition $R \gg \ell_\scat$, necessary for the isotropization of the flux, and for the approximate boundary conditions, is \textit{a priori} not relaxed, although the RTE in itself is also valid in the ballistic regime.

However, we find in the astrophysics literature (radiative transfer in stellar or planetary atmospheres) that the Eddington approximation is very good for isotropic scattering and extends to the optically thin regime~\cite{Unno:1966,Wiscombe:1977}. Nevertheless, the boundary conditions are not discussed and, to our knowledge, there is no other method to treat the boundary conditions within the Eddington approximation. Following Letokhov \textit{et al.}~\cite{Lavrinovich:1975,Johansson:2007,Letokhov:2009}, we have used the method usually applied with the diffusion equation. It is known that these approximate boundary conditions lead to an extrapolation length proportional to the scattering mean-free path~\cite{Rossum:1999,Ishimaru:book,note_extrapolation_length}. In the limit of vanishing scattering, this extrapolation length goes to infinity, and so does the effective size of the medium. This may explain the appearance of a finite random laser threshold in the ballistic limit of the RTE (Eqs.~\ref{eq.ballistic3D},\ref{eq.ballistic2D}).

Finding a better way to treat the boundary conditions, and even including the partial reflection due to the index mismatch, as can be done with the diffusion equation~\cite{Rossum:1999,Ishimaru:book,Wang:book,Zhu:1991,Freund:1992,Aronson:1995}, would certainly improve the validity range and the precision of the RTE threshold.

\section*{APPENDIX C: Partial-wave calculation of lasing thresholds}

This Appendix describes the numerical method used to calculate the
laser threshold of a 2D disordered system in Sec.~\ref{sec.SALT}.
It relies on basis functions that are purely-outgoing at infinity,
called ``constant flux'' (CF) states~\cite{Tureci:2006}.  CF states were originally
introduced in the context of Steady-state Ab-initio Laser Theory
(SALT)~\cite{Tureci:2006,Tureci:2008,Ge:2008,Ge:2010,Cerjan:2015}, a method for accurately calculating above-threshold lasing
solutions. In this work, however, we will not draw upon the full
machinery of SALT, since our interest lies in threshold statistics.
The CF states we shall use are solutions to the wave equation
(\ref{wave equation}), assuming (i) there are \textit{no scatterers},
and (ii) the solutions are purely-outgoing in the external region $r >
R$.  These wavefunctions have the form
\begin{equation}
    u_{mp}(r,\phi) = \left\{
  \begin{array}{ll}
    A_{mp} \, J_m(q_{mp} r) \, \Theta_m(\phi),& r \le R \\
    B_{mp} \, H_m^+(\omega r/c) \, \Theta_m(\phi),& r \ge R,
  \end{array}
  \right.
  \label{ump}
\end{equation}
where $(r,\phi)$ are polar coordinates, $(m,p)$ are azimuthal and
radial quantum numbers, $H_m^+$ denotes Hankel functions of the first
kind, and $\Theta_m(\phi)$ are azimuthal basis functions defined by
\begin{equation}
  \Theta_m(\phi) = \frac{1}{2\pi} \left\{ \begin{array}{ll}
    \sqrt{2} \, \sin\phi, & m > 0\\
    1, & m = 0 \\
    \sqrt{2} \, \cos\phi, & m < 0,
  \end{array}\right.
\end{equation}
which satisfy $\int_0^{2\pi} d\phi\; \Theta_m(\phi)\Theta_{m'}(\phi) =
\delta_{mm'}$.  Matching the wavefunction and its first radial
derivative at $r = R$ gives
\begin{equation}
  \frac{q_{mp}\, J_m'(q_{mp}R)}{J_m(q_{mp}R)} =
  \frac{(\omega/c)\, H_m^{+'}(\omega R/c)}{H_m^+(\omega R/c)},
  \label{qmp}
\end{equation}
which can be solved numerically to find a discrete set of $q_{mp}$
values, corresponding to the different CF states.  With appropriate
normalization (choice of $A_{mp}$), the CF states come to satisfy a
self-orthogonality condition
\begin{equation}
  \int_{r < R} d^2r\; u_{mp} \; u_{m'p'} = \delta_{mm'}^{pp'}.
  \label{normalization}
\end{equation}
Note that the CF basis depends implicitly on the frequency $\omega$,
which appears in Eqs.~(\ref{ump}) and (\ref{qmp}).

We now consider the disordered system with $\epsilon(\bm{r})$ given
by Eq.~(\ref{epsn}).  Its modes can be expanded using the CF basis
states:
\begin{equation}
  \psi(\bm{r}) = \sum_{mp} c_{mp} \, u_{mp}(\bm{r}).
\end{equation}
Such a superposition automatically satisfies outgoing boundary
conditions (with frequency $\omega$), as required of lasing modes.
Plugging this into Eqs.~(\ref{wave equation})--(\ref{epsn}), and using
Eq.~(\ref{normalization}), gives
\begin{multline}
  \sum_{m'p'} \left\{\left[\left(\frac{q_{mp}}{\omega/c}\right)^2 - 1\right]
  \delta_{mm'}^{pp'} - a \sum_j u_{mp}(r_j) \, u_{m'p'}(r_j)
  \right\} c_{m'p'} \\= \chi_g\, c_{mp}.
  \label{eigenval}
\end{multline}
This is a non-Hermitian eigenproblem, whose eigenvalues are the
complex susceptibilities $\chi_g$ that would allow the disordered
structure to lase at frequency $\omega$.  Note that the delta-function
scatterers enter in the second term in the matrix; their
delta-function nature is handled ``exactly'', in the sense that we
need not approximate them through spatial discretization.

In order to solve the eigenproblem numerically, we truncate to a
finite CF basis set.  For $\omega = 60 c/R$ (see Section
\ref{sec.SALT}), we take $m \le 75$ and $\mathrm{Re}(q_{mp}) \le
180/R$.  Essentially, these truncations limit the resolution of the
wavefunction in the azimuthal and radial directions, respectively.
There are 6098 CF states in the remaining basis set.  The matrix in
Eq.~(\ref{eigenval}) is non-sparse, so the solution time increases
with the basis size, $M$, as $O(M^3)$.

\begin{figure}[t]
\centering
\includegraphics{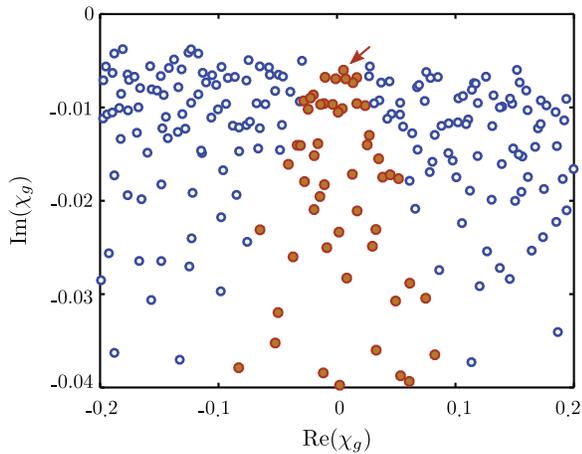}
\caption{Numerical values of $\chi_g$ obtained for a disorder
  realization with $N = 250$ scatterers and $\ell_\scat = 0.15 R$,
  with $\omega = 60 c/R$.  The susceptibility eigenvalues are truncated
  to those with sufficiently small real parts (solid circles), then
  the eigenvalue with smallest $|\mathrm{Im}(\chi_g)|$ (indicated by
  an arrow) determines the laser threshold. }
\label{fig.chig}
\end{figure}

Fig.~\ref{fig.chig} shows the computed values of $\chi_g$ for a
typical disorder realization.  The eigenvalues with very large
$\mathrm{Re}[\chi_g]$ are not the lasing modes we are interested in;
those are modes confined because of a large real uniform background
susceptibility $\chi_g$, rather than random scattering.  We filter out
these solutions by truncating the eigenvalues to those with
sufficiently small real parts (specifically, $|\mathrm{Re}[1/\chi_g]|
< 3 |\mathrm{Im}[1/\chi_g]|$).  These remaining eigenvalues form a
random distribution in $\mathrm{Im}[\chi_g]$, i.e.,~the amplification
provided by the gain medium.  Their residual small but non-zero
$\mathrm{Re}[\chi_g]$ correspond to the index shifts necessary to make
each mode lase at frequency $\omega$. Varying $\omega$ moves these
eigenvalues mostly sideways in the complex plane, without much change
in $\mathrm{Im}[\chi_g]$. As described in Section \ref{sec.SALT}, we
then pick the smallest eigenvalue with the smallest value of
$|\mathrm{Im}[\chi_g]|$, which determines the gain length $\ell_g$.
Shifts in the real part of $\chi_g$ due to the gain may be relevant for the study of individual laser thresholds, but are neglected here as they do not influence the overall statistics of laser threshold we are interested in.



\end{document}